# Mechanism of Electric Field Induced Conductance Transition in Molecular Organic Semiconductor Based Thin Films

**Ruchi Agrawal** and **Subhasis Ghosh**
School of Physical Sciences
Jawaharlal Nehru University, New Delhi 110067

## Abstract

We have studied the electrical field induced conductance transition in thin film of Perylenetetracarboxylic dianhydride sandwiched between two metal electrodes, from an insulating state to conducting state with a high ON-OFF ratio in those devices, where one of electrodes is either Al or Cu. Temperature dependence of resistivity shows semiconducting behavior in OFF-state, but it shows metallic behavior in the ON-state. Devices with a thin intermediate layer of LiF between metal electrode and organic layer, or devices fabricated in planar configuration don't show switching behavior. All these suggest that conducting pathways are responsible for the electric field induced conductance transition.

There is increasing interest in solid-state memory in two terminal devices based on organic semiconductor. In particular, the two terminal nonvolatile memory devices based on organic semiconductor sandwiched between two metal electrodes have attracted much attention for their advantage due to low power consumption, relatively simple fabrication technique and absence of technological bottleneck on miniaturization. Since the discovery[1] of bistable switching in organometallic charge transfer complex, there are several demonstrations of bistable switching in different types of organic and inorganic thin films. The electric field induced bistability of the resistance/conductance state has been observed in inorganic oxide[2,3], colossal magnetoresistance[4], and organic materials[5-11]. We have recently reported[12] similar switching behavior in 8-Hydroxyquinoline Aluminum(Alq3) based two terminal memory devices. Though several mechanisms have been proposed to explain the bistable phenomenon in various materials with imperfect structures, but it seems the nanoparticles or nano-filamentary metallic pathways play an important role on the electrical bistability in these systems. Recently, it has been demonstrated[11,13] that metal nanoparticles embedded in the organic matrix showed similar switching behavior. The UCLA group has shown[14] that migration of Cu ions in the organic thin film plays dominant role on this kind of conductance transition.

In this letter, we report bistable electrical switch, based on electric field induced conductance transition between insulating and conducting states, giving rise to two distinct states: (i) ON-state that allows the charge transfer and (ii) OFF-state that blocks the charge transfer, in an important organic molecular semiconductor,3, 4,9,10 Perylenetetracarboxylic dianhydride(PTCDA). During the last decade, organic field-effect transistors and light-emitting diodes based on PTCDA have been demonstrated. PTCDA is a planar molecule, grown on a large variety of substrates in a polycrystalline film morphology with the molecular planes oriented nearly parallel to the substrate surface[15]. Hence, a seamless integration of electrical and optical devices for multifunctionality on a single device may be achieved using PTCDA molecules.

The devices are fabricated in two configurations. In sandwiched configuration, single layer of PTCDA evaporated between top and bottom electrodes. In planar configuration, two metal electrodes separated by 5-10μm are deposited on single layer of PTCDA, which is grown either on glass, or thermally grown $SiO_2$ on Si. In some sandwiched devices a thin(~3-5nm) layer of LiF was introduced between top metal electrode and PTCDA layer. High purity sublimed PTCDA have been procured from Sigma Aldrich chemical co. and evaporated on different metal electrodes, Al, Cu, Au and ITO(indium tin oxide). All evaporations were done at a base pressure of $10^{-6}$ Torr at a rate of 1-2 Å/Sec. Thermally evaporated PTCDA layer were characterized by atomic force microscopy and absorption spectroscopy.

The current-voltage(J-V) characteristics are studied in metal/PTCDA/metal devices by varying the thickness of the organic layer as well as by using different electrodes. Fig.1 shows the room temperature J-V characteristics of Al/PTCDA/Al and Cu/PTCDA/Cu structures. The J-V characteristics were reproducible in the low bias region up to a threshold value. At this threshold voltage($V_T$), current increases sharply by several order of magnitude and the device undergoes a conductance transition from an insulating state to a conducting state. In second bias scan, the device remains in conducting state by displaying high current even at low bias below the $V_T$ which persists in subsequent bias scans. In the conducting state, current shows the ohmic behavior($J \propto V$). Fig.2 shows the field induced switching behavior when one of the electrodes replaced by ITO. Similar switching behavior is also observed in Au/PTCDA/Al and Au/PTCDA/Cu devices, but no switching is observed when both the electrodes are Au i.e. in Au/PTCDA/Au and Au/PTCDA/ITO structures. In summary, this type of switching behavior is always observed in sandwiched devices (i) when both the electrodes are Al or Cu, and (ii) when one of the electrodes is either Al or Cu. Nonconducting state shows high impedance of the order of 1GΩ and conducting state shows low impedance of order of 1MΩ. The device remains in conducting state for several hours to days. The nonconducting or insulating state can be recovered by keeping the sample in zero bias at high temperature (~350-400K) or by applying negative bias pulse at room temperature. $V_T$ varies from 1V to 37V and ON-OFF ratio varies from $10^2$–$10^4$ depending on the thickness of the PTCDA layers and the type of electrodes. Highest ON-OFF ratio(~$10^4$) is observed in devices when both the electrodes are Al and thickness of the PTCDA layer is 200nm.

There are several scenarios proposed for this electric field induced switching. One scenario is the electric field induced conformational changes in organic molecules, but there is no direct evidences in terms of spectroscopic signature of different conformation of organic molecules when device is in conducting or insulating states. This scenario may be relevant[7,16,17] for switching based on single organic molecule. Another scenario is based on the formation of metal filament in molecular thin film. To reveal the mechanism, we have measured the temperature dependence of the metal/PTCDA/metal devices in the conducting state and insulating state. It is clear from Fig.3 that sample shows insulating behavior(resistivity decreases with temperature) when the device is in OFF-state and weakly metallic behavior(resistivity increases with temperature) when the device is in ON-state. PTCDA is an electron transport organic semiconductor with ionization potential and electron affinity of 6.4eV and 4.2eV[18], respectively, so efficient electron injection is possible when a low work function metal like Al is used as cathode. An activation energy of 0.17eV, which is barrier for electron injection from metal to PTCDA at Al/PTCDA[18] interface, has been found from the temperature dependence of resistance of PTCDA thin film in the insulating state. But, PTCDA is an organic semiconductor with a band gap of 2.2eV[18]. Hence, the metallic behavior in the ON-state is not expected. The metallic conduction can be explained either by the percolation among the metal nanoparticles or by the impurity band conduction[10]. The latter mechanism should lead to negative differential resistance and N-shaped J-V characteristics, which has not been observed in our devices. It has been attempted[19,20] to explain the sharp increase of current above $V_T$ by tunneling between the metal nanoparticles, but Ohmic behavior in the ON-state does not support this model. Previous studies[21,22] have shown that diffusion barrier for Cu and Al is very small in semiconducting materials and it is also known[14] that metal atoms can migrate inside organic layer during metallization and can form metal clusters or metal nanoparticles. These results indicate that the migration of Al or Cu atoms and subsequent aggregation to metal cluster or nanoparticles in organic layer during evaporation of top electrode plays important role on the bistable effect. And, the metallic behavior in the ON-state can be understood as electric field induced percolation through a conducting pathways, like metallic wire, consisting of metallic nanoparticles. The resistance of this conducting channel has been found to be 45KΩ at low temperature by extrapolating the linear region of the curve in Fig.3(b). Similar high resistance has been observed[23] in one dimensional metallic wire. Using secondary ion mass spectroscopy, it has been shown that when the concentration of Cu ions, migrated from metal electrode is high, the device shows conducting state and when the concentration is low, the device shows insulating state. Essentially, metal ions are driven out and into the organic layer in insulating and conducting states, respectively. When a bias more than $V_T$ is applied, metal ions drift inside the organic solid and finally reach cathode electrode giving rise to metallic pathway responsible for ON-state. When the bias is switched off metal nanoparticles relaxed and the device switches back to OFF-state with a time constant, which is called retention time. We have observed that retention time in our devices can be as long as few days and depends on temperature, but details dynamics of retention time has not been studied. An important point to mention here that the devices are not shorted in the ON-state. It is clear from Fig.3 that the relative change of resistance of the device between 80K and 300K is 1.2, whereas that in case of bulk Al is around 6. We have not observed similar switching behavior in case of other organic semiconductor, like metal phthalocyanines sandwiched between Al or Cu electrodes, but it can be observed reproducibly in Alq3 and PTCDA, which are strongly correlated organic semiconductors[22,24]. It will be interesting to investigate whether strong electron-electron interaction plays any role, which has been discussed in a recent paper[25].

Further support for percolative conduction among metal nanoparticles can be provided by the results given in Fig.4, which shows the J-V characteristics for two subsequent bias scans in (i) planar devices and (ii) sandwiched devices in which a thin layer of LiF is introduced between top electrode and PTCDA layer. It is very unlikely that migration of metal atoms from electrode is possible in these two cases. As expected, switching behavior is not observed in these devices. Typically ON-OFF transitions are induced by an electric field of 60-100KV/cm which is required for field induced percolation. It is clear in Fig.4 that similar electric field(~100KV/cm) cannot induce conductance transition, because in this planar devices it is not possible to have continuous metallic pathways between planar electrodes.

In conclusion, we report a organic semiconductor based bistable switch using conductance transition in single layer devices based on PTCDA sandwiched between two electrodes. Switching has been observed in those devices in which both or one of the electrodes is either Al or Cu. The mechanism is explained by the migration of metal atoms from metal electrodes to organic layer during the top electrode evaporation. High retention time of both ON and OFF states is promising for memory applications.

One of the authors(RA) thanks to UGC-CSIR for the financial support through fellowship. This work was partly supported by Department of Science and Technology, Government of India.

**Figure Captions**

Figure 1. *J-V* characteristics of single layer devices based on PTCDA thin film, sandwiched between two metal electrodes. At room temperature, current increases by almost four order of magnitude at a threshold bias($V_T$) of 1.5V in (a) Al/PTCDA(200nm)/Al and by almost two order of magnitude at $V_T$ of 1.2V in (b) Cu/PTCDA(200nm)/Cu. Empty circles with connecting lines (a guide to eye) represent the first bias scan and empty squares for second bias scan. In all devices, after crossing the $V_T$, the devices remain in the conducting state which is observed in second, third, and subsequent bias scans.

Figure 2. *J-V* characteristics of single layer devices based on PTCDA thin film, sandwiched between two metal electrodes. In this case one of the electrodes is ITO instead of Al. At room temperature, current increases by almost two order of magnitude at a threshold bias($V_T$) of 12V in (a) Cu/PTCDA(400nm)/ITO and 37V in (b) Al/PTCDA(400nm)/ITO. Empty circles with connecting lines (a guide to eye) represent the first bias scan and empty squares for second bias scan. In all devices, after crossing the $V_T$, the devices remain in the conducting state which is observed in second, third, and subsequent bias scans.

Figure 3. Temperature dependence of resistance of PTCDA in Al/PTCDA(200nm)/Al device in (a) insulating state and (b) conducting state. Inset shows the Arrhenious plot of resistance vs. temperature in insulating state. The solid line fit gives an activation energy of 0.17eV.

Figure 4. *J-V* characteristics of single layer devices in, (a) sandwiched configuration between two Al electrodes with a thin(~4nm) LiF layer between top electrode and PTCDA layer, (b) planar configuration with two Al electrodes separated by 10μm. Empty circles and empty squares represent the first bias scan and second bias scan, respectively. In these two cases, no switching has been observed.

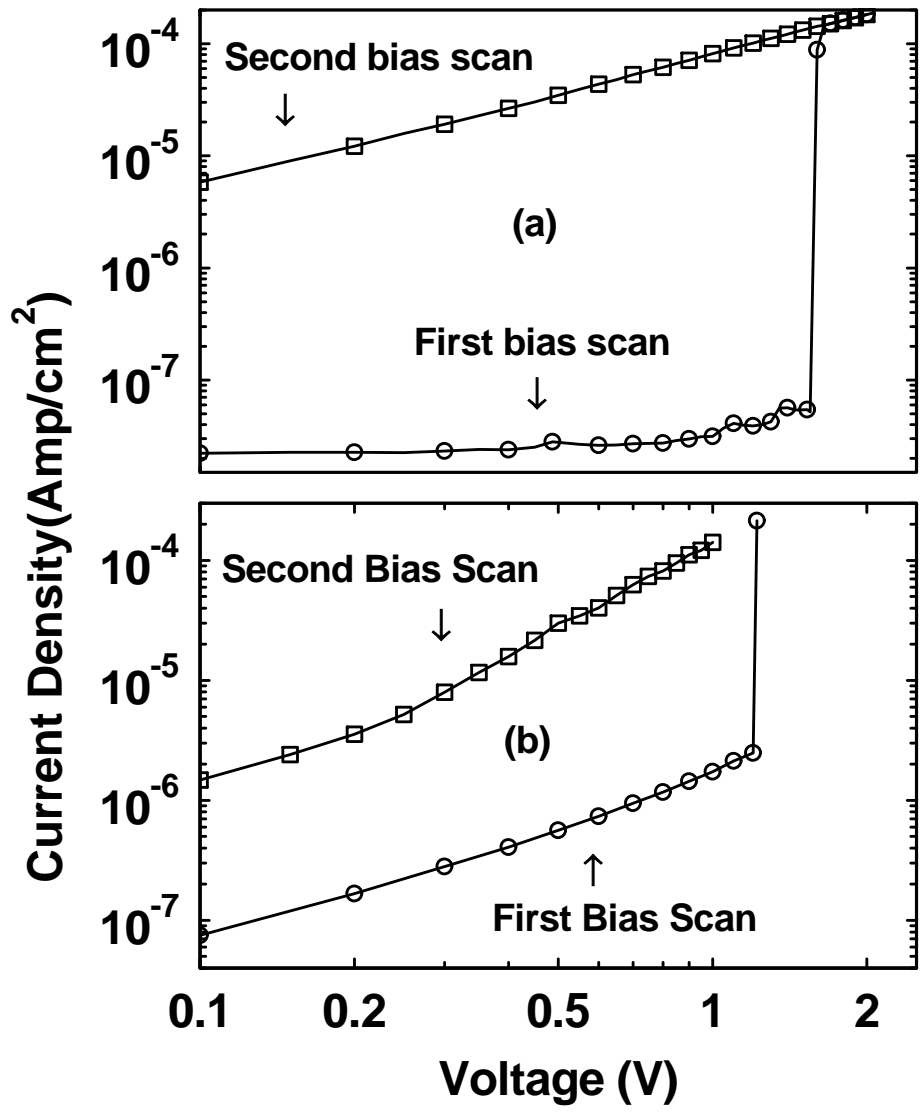

**Figure 1**

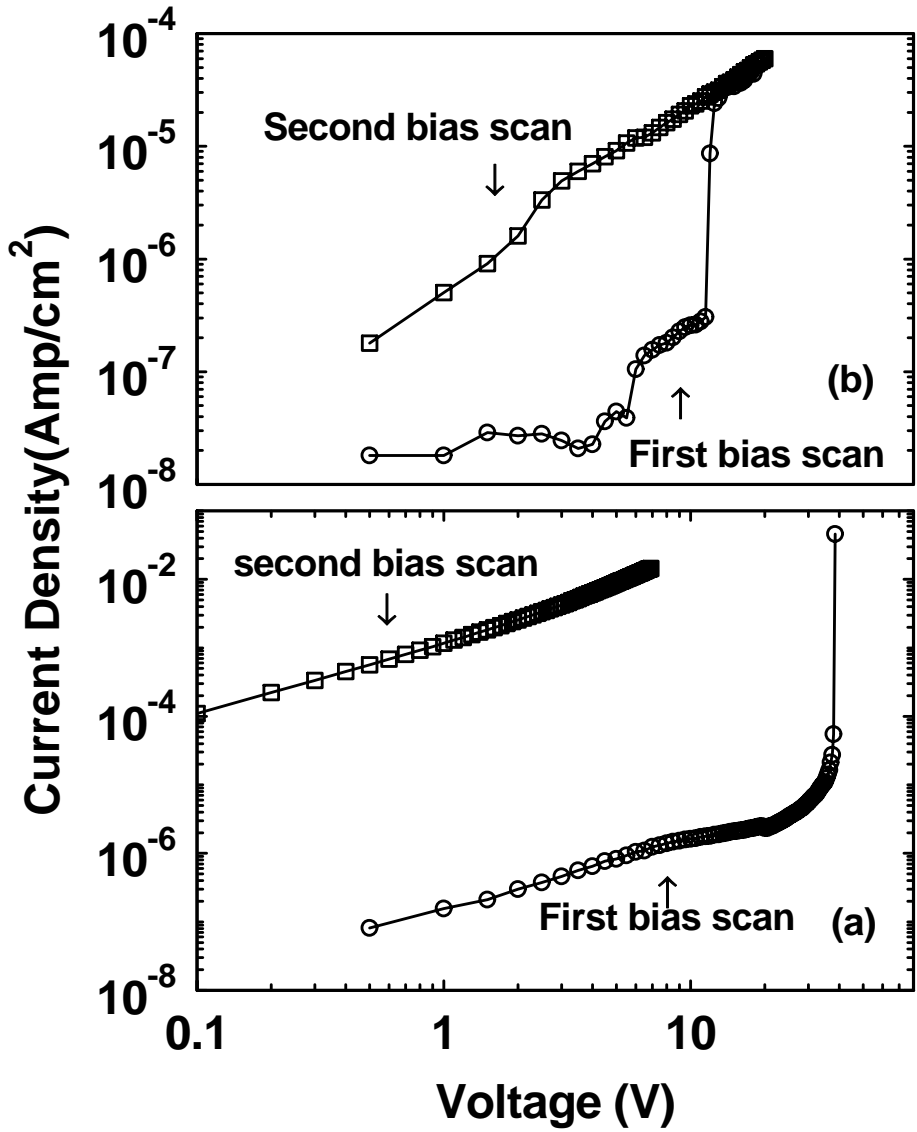

**Figure 2**

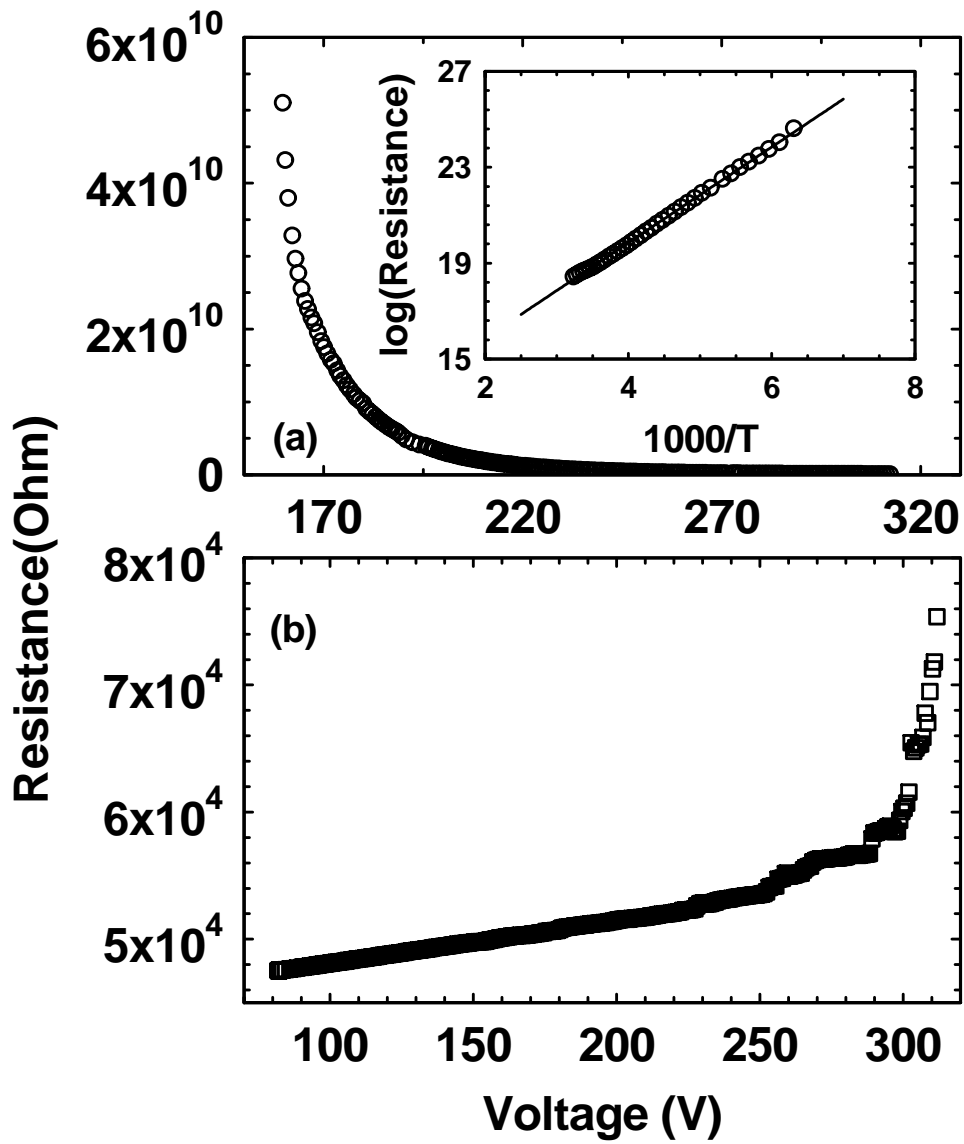

Figure 3

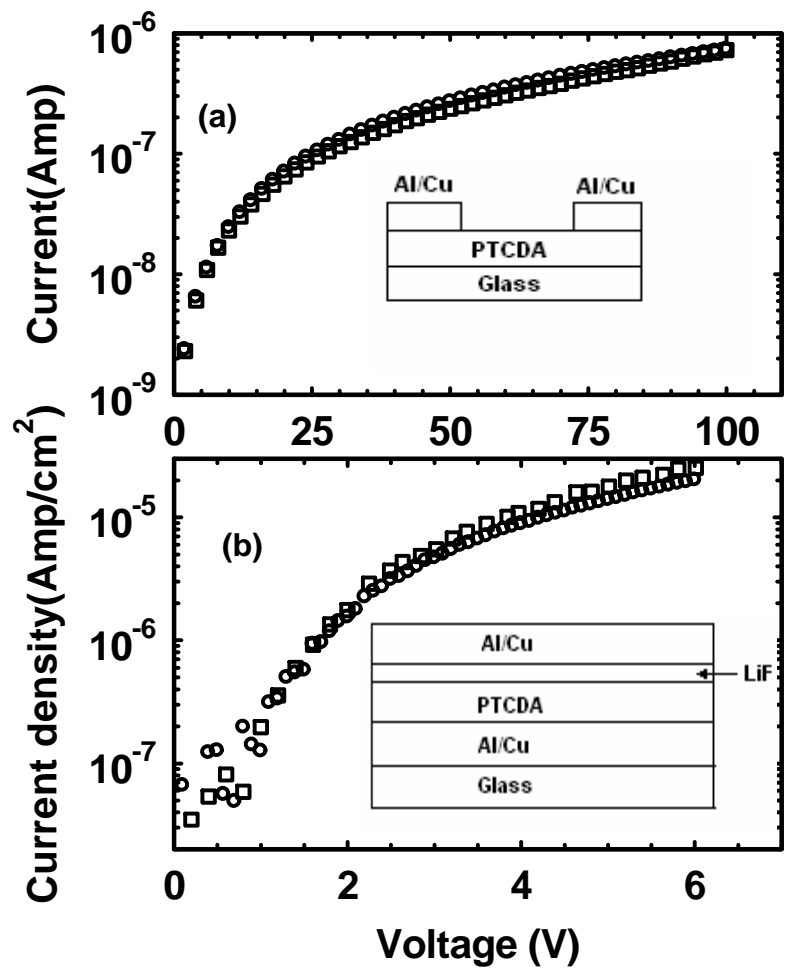

**Figure 4**